\documentclass[a4paper,11pt]{article}
\usepackage{pos}

\usepackage{comment}

\newcommand{\ket}[1]{\ensuremath{| {#1} \rangle }}
\newcommand{\bra}[1]{\ensuremath{\langle {#1} |}}

\renewcommand{\d}{\mathrm{d}}

\title{The smeared $R$-ratio in isoQCD from first-principles lattice simulations}

\author*[a]{Francesca Margari} 

\author[b]{Simone Bacchio}
\author[c]{Alessandro De~Santis} 
\author[d]{Antonio Evangelista}
\author[a]{Roberto Frezzotti} 
\author[e]{Giuseppe Gagliardi}
\author[f]{Marco Garofalo}
\author[g]{Francesco Sanfilippo}  
\author[a]{Nazario Tantalo}

\onbehalf{for the Extended Twisted Mass collaboration}

\affiliation[a]{Dipartimento di Fisica and INFN, Universit\`a di Roma ``Tor Vergata",\\ 
Via della Ricerca Scientifica 1, I-00133 Rome, Italy}

\affiliation[b]{Computation-based Science and Technology Research Center, The Cyprus Institute,\\20 Konstantinou Kavafi Street, 2121 Nicosia, Cyprus}

\affiliation[c]{Helmholtz-Institut Mainz, Johannes Gutenberg-Universität Mainz, 55099 Mainz, Germany \\ GSI Helmholtz Centre for Heavy Ion Research, 64291 Darmstadt, Germany}

\affiliation[d]{Department of Physics, University of Cyprus, P.O. Box 20537, 1678 Nicosia, Cyprus}

\affiliation[e]{Dipartimento di Matematica e Fisica, Universit\`a Roma Tre and INFN, Sezione di Roma Tre,\\Via della Vasca Navale 84, I-00146 Rome, Italy}

\affiliation[f]{HISKP (Theory), Rheinische Friedrich-Wilhelms-Universit\"at Bonn,\\Nussallee 14-16, 53115 Bonn, Germany}

\affiliation[g]{Istituto Nazionale di Fisica Nucleare, Sezione di Roma Tre,\\Via della Vasca Navale 84, I-00146 Rome, Italy}

\emailAdd{francesca.margari@roma2.infn.it}

\abstract{The $R$-ratio is a phenomenological observable of great relevance, both in itself and in applications such as the dispersive approach to the muon anomalous magnetic moment. It can be investigated from first-principles with controlled statistical and systematic errors in lattice QCD by introducing an arbitrary smearing kernel and employing spectral reconstruction techniques, such as the well-known Hansen-Lupo-Tantalo method. Improving upon a first study published in 2023, we show preliminary results using the correlation functions produced by ETMC in $N_f = 2+1+1$ lattice simulations at four lattice spacings, different volumes and with higher statistics w.r.t. our previous study. The new correlators, thanks to the implementation of the Low Mode Average technique, allow the determination of the $R$-ratio smeared with Gaussian kernels of widths down to $\sigma \sim 200$ with phenomenologically relevant precision.}

\FullConference{The 42nd International Symposium on Lattice Field Theory (LATTICE2025)\\
2-8 November 2025\\
Tata Institute of Fundamental Research, Mumbai, India\\}

\begin{document}
\maketitle

\section{Introduction}
\label{sec:introduction}
The $R$-ratio between the $e^+e^-$ cross-section into hadrons and that into muons plays a central r\^ole in elementary particle physics since its introduction in~\cite{Cabibbo:1970mh}. In recent years, its importance has been mostly associated with the dispersive determination of the leading hadronic vacuum polarization (HVP) contribution to the muon anomalous magnetic moment, $a_\mu$.

\begin{figure}[h]
\begin{center}
\begin{minipage}[c]{0.55\textwidth} % [c] = center vertical alignment
    \centering
    \includegraphics[height=0.15\textheight]{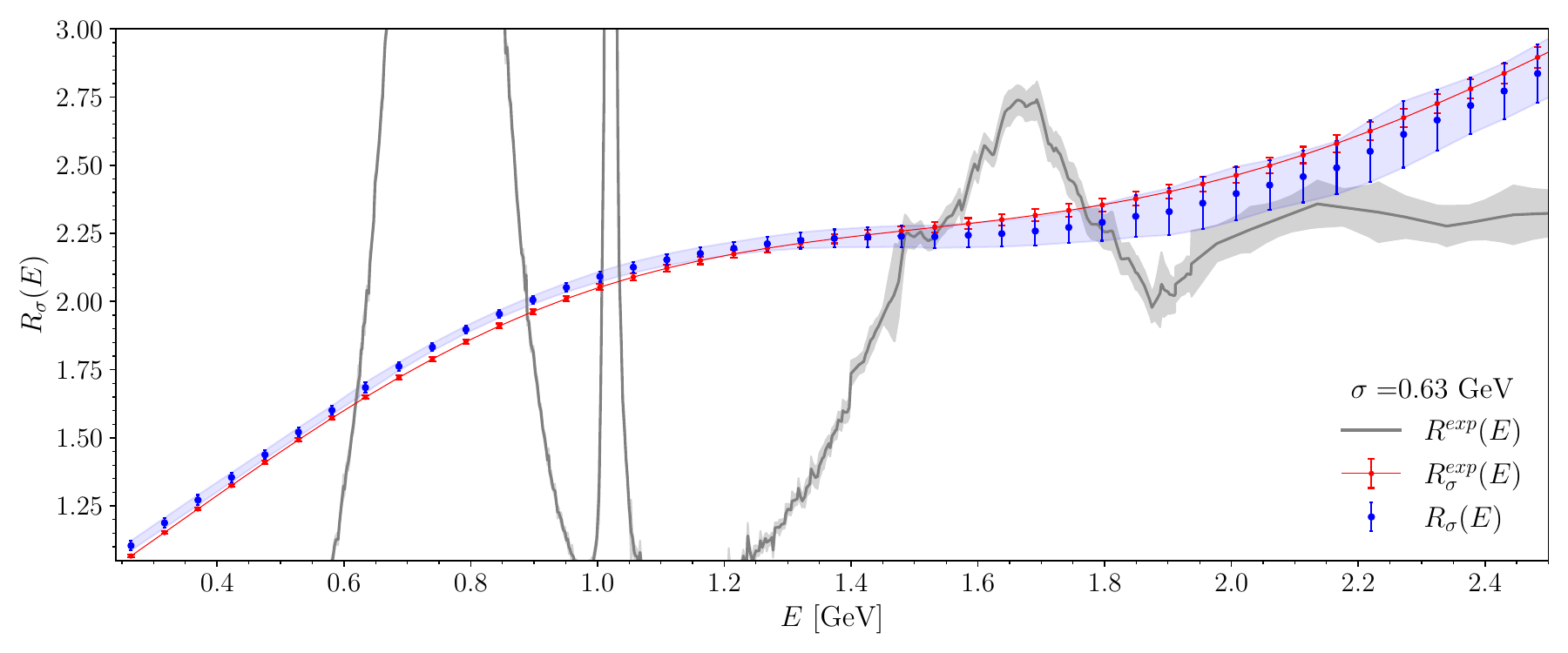}\\[0.5em]
    \includegraphics[height=0.15\textheight]{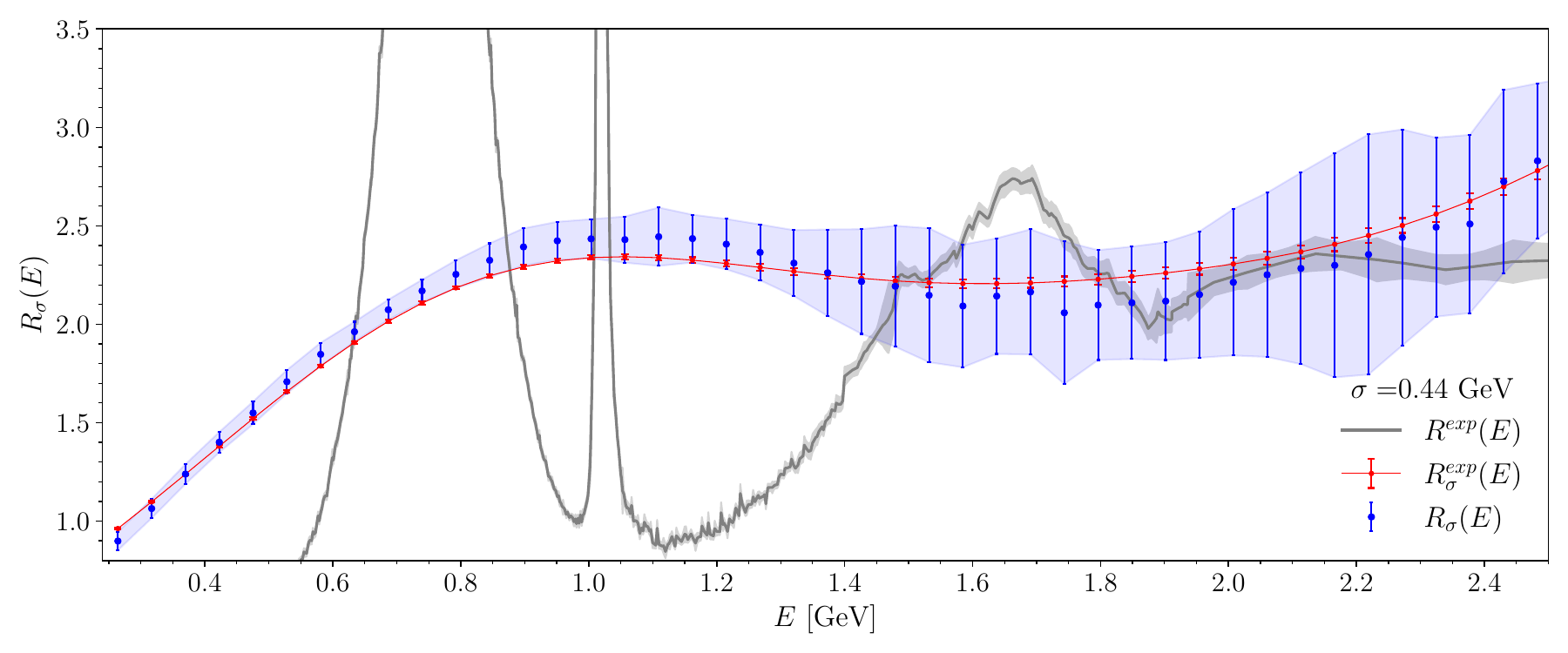}
\end{minipage}%
\hfill
\begin{minipage}[c]{0.44\textwidth} % [c] = center vertical alignment
    \centering
    \includegraphics[height=0.15\textheight]{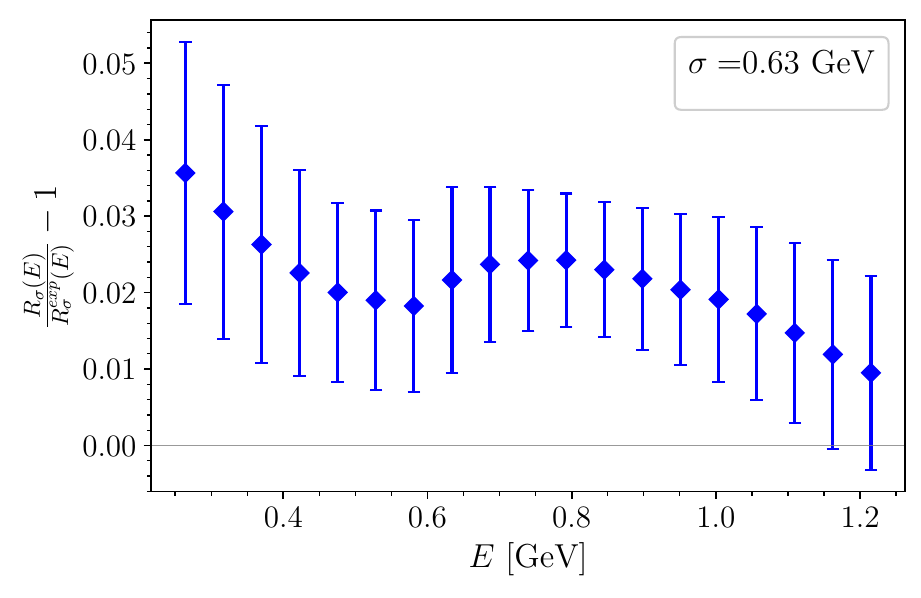}\\[0.5em]
    \includegraphics[height=0.15\textheight]{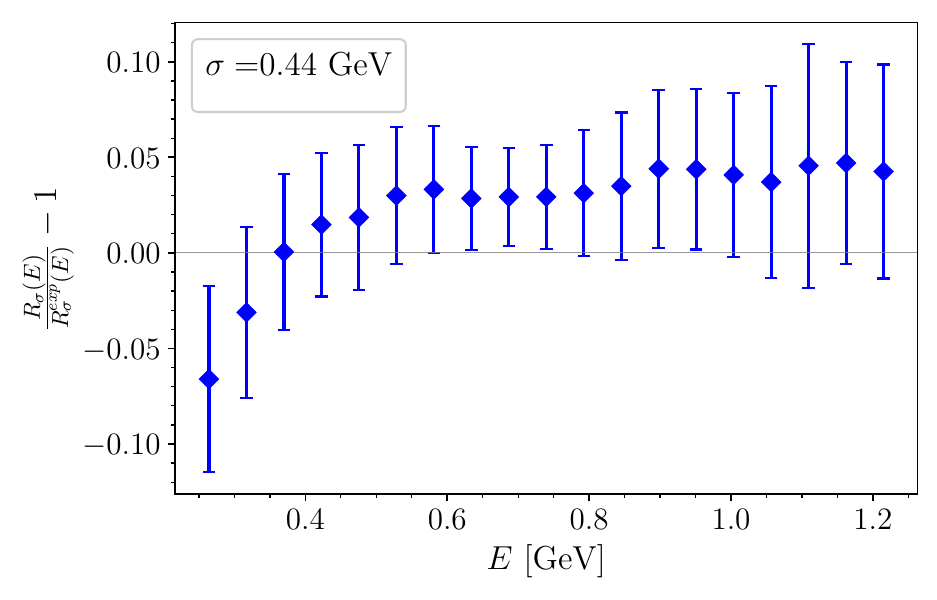}
\end{minipage}
\caption{Comparison of $R_\sigma(E)$ (blue points) and $R^\mathrm{exp}_\sigma(E)$ (red points) as functions of $E$ for $\sigma=0.63$~GeV and $0.44$~GeV. The relative difference $R_\sigma(E)/R^{\mathrm{exp}}_\sigma (E)-1$ is also shown as a function of the energy. For $\sigma \simeq 0.6$~GeV, the total relative uncertainty is at the level of $\sim 1\%$ and a deviation of about $3\sigma$ with respect to the experimental $e^+e^-$ data can be observed. For $\sigma \simeq 0.4$~GeV, the total relative uncertainty increases to $\sim 3.5\%$, preventing us from resolving statistically significant deviations from the experimental determination.}
\label{fig:PRL-Rratio}
\end{center}
\end{figure}

In Ref.~\cite{ExtendedTwistedMassCollaborationETMC:2022sta}, a first-principles lattice QCD investigation of the $R$-ratio smeared with Gaussian kernels is addressed using the Hansen-Lupo-Tantalo (HLT) spectral-density reconstruction method~\cite{Hansen:2019idp}.
Specifically, the $R$-ratio is convoluted with Gaussian smearing kernels, with widths ranging from 440 MeV to 630 MeV and center energies up to 2.5 GeV. The smeared $R$-ratio is extracted according to 
\begin{flalign}
R_\sigma(E) = \int_{0}^{\infty} d \omega \; G_\sigma (E-\omega) R(\omega) \; , 
\label{eq:rsigma_1}
\end{flalign}
with normalized Gaussian kernels, $G_\sigma(E-\omega)= \exp{(-(E-\omega)^2/2 \sigma^2)/\sqrt{2\pi\sigma^2}} $.
In Fig.~\ref{fig:PRL-Rratio}, the lattice results of the smeared $R$-ratio obtained in~\cite{ExtendedTwistedMassCollaborationETMC:2022sta}, is compared with the corresponding quantity derived from the KNT19 compilation~\cite{Keshavarzi:2019abf} of $ R$-ratio experimental data smeared with the same Gaussian kernel. For central energies of the smearing Gaussian in the region around the $\rho$ resonance the results were sufficiently
precise to let us observe a tension of about three standard
deviations with experiments. 

Building upon this first study, in this work we present updated preliminary results for the smeared $R$-ratio based on ETMC gauge ensembles with four lattice spacings, different volumes, and significantly higher statistics. We employ the Low Mode Averaging (LMA) technique, which has proven to be highly beneficial for improving the signal-to-noise ratio at large Euclidean times. This progress enables the determination of the $R$-ratio smeared with Gaussian kernels of widths as small as $\sigma \sim 200$ MeV with phenomenologically relevant precision, allowing for a clear resolution of the $\rho$ resonance around 770~MeV.

\section{Methods}
\label{sec:methods}
We compute the two-point Euclidean correlator of the quark electromagnetic current given by
\begin{flalign}\label{eq:corr}
C(t)= - \frac{1}{3}\sum_{i=1}^3 \int \d^3x \mathrm{T} \bra{0} J_i(x) J_i(0) \ket{0} \; ,
\end{flalign}
where $J_\mu=\sum_{f}q_f \bar \psi_f \gamma_\mu \psi_f$ is the electromagnetic current, with $f=\{u,d,s,c,b,t\}$, $q_{u,c,t}=2/3$ and $q_{d,s,b}=-1/3$. We consider two lattice discretizations of the e.m. currents $J^{\mathrm{reg}}_{\mu, f}$ with reg = $\{$TM, OS$\}$ corresponding
to the so-called Twisted-Mass (TM) and Osterwalder-Seiler (OS) regularizations, see Ref.~\cite{ExtendedTwistedMassCollaborationETMC:2024xdf} for details. 

In order to improve the signal-to-noise ratio of correlation functions at large distances, we employ the Low Mode Averaging (LMA) technique, which is widely used in lattice QCD as a noise reduction method (see~\cite{Neff, Giusti:2004yp, DeGrand:2004qw, Borsanyi:2020mff} for details). As explained in~\cite{Margari:20253h}, this approach isolates the contribution of the low eigenmodes of the Dirac operator from the rest of the spectrum. The corresponding eigenvectors are computed exactly and used to construct all-to-all correlation functions, allowing one to extract complete information from each gauge configuration. When a sufficiently large number of modes is included, the correlation function at large distances is dominated by the low-mode contribution, leading to a substantial reduction of statistical uncertainties.

The vector-vector two-point correlation function is our primary observable, directly computed via lattice simulations, and it is connected to the $R$-ratio in the so-called time momentum representation \cite{Bernecker:2011gh}
\begin{flalign}
C(t)= \frac{1}{12 \pi^2} \int_{E_{\mathrm{th}}}^{\infty}\d \omega\, e^{-\omega t} \;\omega^2 R(\omega) \; ,
\end{flalign}
where the threshold energy $E_{\mathrm{th}}$ is $2m_\pi$ in isoQCD. Theoretically $R(\omega)$ is a distribution, the spectral density of the correlator $C(t)$, and can conveniently be probed by using suitable smearing kernels. In order to compute $R_\sigma(E)$ on the lattice, see Eq.~(\ref{eq:rsigma_1}), we employ the HLT method. According to this approach, the smearing kernels are approximated as 
\begin{flalign}
K(\omega; \mathbf{g}) = \sum_{\tau=1}^{\tau_{\rm max}} g_{\tau} e^{-a \omega \tau} \; ,
\end{flalign}
where $\tau$ is an integer variable and $a$ is the lattice spacing. The distance between the target kernel and its representations in terms of the coefficients $g_{\tau}$ is measured by the functionals
\begin{flalign}
A_{\rm n} [\mathbf{g}] = \int_{E_0}^{\infty} \d \omega  \ w_{\mathrm{n}}(\omega) \left| K(\omega; \mathbf{g}) - \frac{12\pi^2G_\sigma(E-\omega)}{\omega^2}  \right|^2 \, , 
\end{flalign}
that, for weight-functions $w_{n}>0$, correspond to a class of weighted $L_2$-norms in functional space. We have considered the following weight functions
\begin{flalign}
	&w_\alpha(\omega) = e^{a\omega \alpha}\;,
	\qquad
	\alpha=\left\{0,\frac{1}{2},2^-\right\}\;,
	\label{eq:walpha}
\end{flalign}
that we distinguish by using the tag $\mathrm{n}=\{0,1,2^-\}$. The \textbf{g} coefficients result from minimizing
\begin{flalign}\label{eq:W-functional}
W_{\rm n} [\lambda, \mathbf{g}] = (1 - \lambda)\frac{A_{\rm n}[\mathbf{g}]}{A_{\rm n}[\mathbf{0}]} + \lambda   B [\mathbf{g}], \quad \text{where} \ B [\mathbf{g}] = \sum_{\tau_{1,2}=1}^{T/a} g_{\tau_1}g_{\tau_2} \mathrm{Cov}_{\tau_1 \tau_2} \; ,
\end{flalign}
where $\lambda$ is a parameter varied/optimized in  stability analysis. For a complete and detailed description, we refer the reader to the Supplementary Material section in Ref.~\cite{ExtendedTwistedMassCollaborationETMC:2022sta}.

\section{Materials}
\label{sec:materials}
The lattice gauge ensembles employed in this study, produced by the ETMC, are summarized in Table~\ref{tab:ensembles} and described in detail in Ref.~\cite{ExtendedTwistedMassCollaborationETMC:2024xdf}. In particular, we exploit the mixed-action setup adopted in Ref.~\cite{Frezzotti:2004wz, windows}. We use both the Twisted Mass (TM) and Osterwalder--Seiler (OS) lattice-regularized correlators $C(t)$ and, in this way, we can achieve a more reliable estimate of the systematic uncertainties associated with the continuum extrapolation. The corresponding determinations of $R_\sigma(E)$ obtained with the two regularizations differ by $O(a^2)$ discretization effects~\cite{Frezzotti:2003ni,Frezzotti:2005gi}, and must coincide within errors in the continuum limit. 
The total contribution $R_\sigma(E)$ can be obtained by considering separately the contributions corresponding to connected (C) and disconnected ($\mathrm{disco}$) fermionic Wick contractions of $C(t)$. The connected part can be further decomposed according to the flavour content of the electromagnetic currents, leading to
\begin{flalign*}
R_\sigma(E) =
R^{\ell \ell, C}_\sigma (E)
+ R^{ss, C}_\sigma (E)
+ R^{cc, C}_\sigma (E)
+ R^{\mathrm{disco}}_\sigma (E)\,.
\end{flalign*}
We considered three different values of $\sigma$,  namely  $
\sigma_1=0.4~\mathrm{GeV}$, $\sigma_2=0.3~\mathrm{GeV}$, $\
\sigma_3=0.2~\mathrm{GeV}$ and central energies $E$ in the range $[0.3,1.2]$~GeV. 
The analysis described in the following has been performed for the light--light connected contribution $R^{\ell \ell, C}_\sigma (E)$ for all ensembles, regularizations, values of energy and $\sigma$. The full statistics data are blinded, further details will be given in Ref.~\cite{ETMC_light}.
\begin{table}[!h]
\centering
\begin{tabular}{lllc}
\hline
ensemble & $(L^3 \times T)/a^4$ & $a$ [fm] & $L$ [fm]  \\
\hline
B64   & $64^3 \times 128$  & 0.0795 & 5.09 \\
B96   & $96^3 \times 196$  & 0.0795 & 7.64  \\
C80   & $80^3 \times 160$  & 0.0682 & 5.46  \\
D96   & $96^3 \times 192$  & 0.0569 & 5.46 \\
E112  & $112^3 \times 224$ & 0.0489 & 5.46 \\
\hline
\end{tabular}
\caption{ETMC ensembles \cite{poster-Bartosz} used for two-point vector correlators in isoQCD.}
\label{tab:ensembles}
\end{table}

\section{Data analysis}
\label{sec:dataanalysis}
\begin{figure}[]
\centering
\includegraphics[width=0.8\textwidth,height=0.3\textheight]{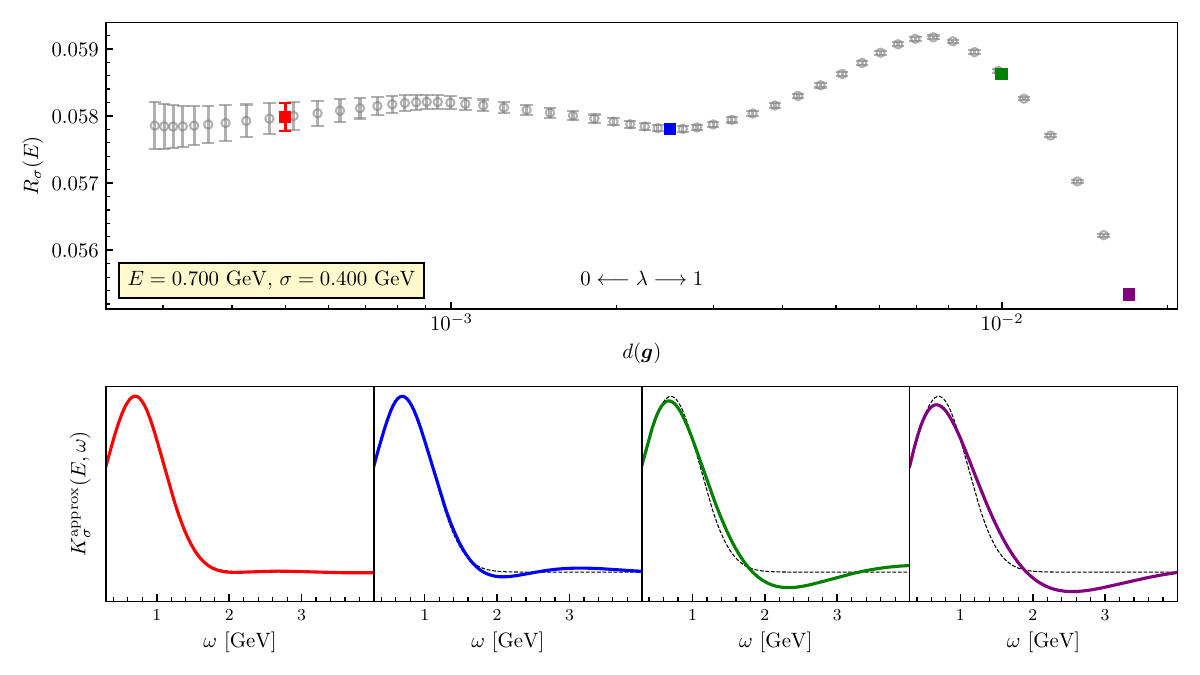}
    \caption{Stability analysis  of $R_\sigma(E)$ for different values of $d(\mathbf{g})$ on the B64 ensemble at energy $E=0.70$ GeV and $\sigma=0.4$ GeV. As $\lambda \to 0$, the reconstructed kernel approaches the target one, at the price of larger statistical uncertainties. Conversely, as $\lambda \to 1$, the statistical precision improves while the reconstructed kernel deviates significantly from the target, reflecting the competition between resolution and statistical uncertainty in Eq.~(\ref{eq:W-functional}). Our best estimate for $R_\sigma(E)$ is selected from this statistically dominated regime, corresponding to small values of $d(\mathbf{g})$ where stability within the statistical errors is observed.
}\label{fig:changelambda}
\end{figure}

\begin{figure}
\centering
\includegraphics[width=0.8\textwidth,height=0.4\textheight]{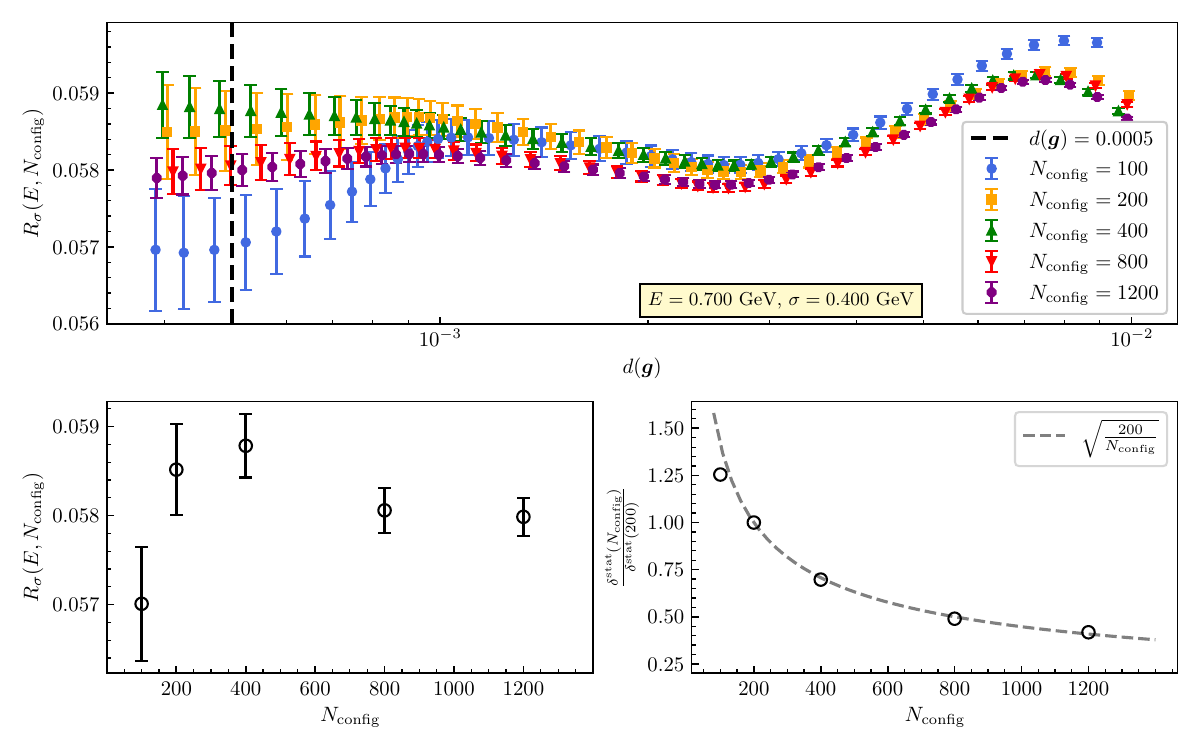}
    \caption{\textit{Top panel:} Stability of $R_\sigma(E)$ as a function of $d(\mathbf{g})$ for different values of $N_{\mathrm{config}} = 100, 200, 400, 800, 1200$, 
    at $E=0.70~\mathrm{GeV}$ and $\sigma=0.4~\mathrm{GeV}$ on the B64 ensemble. \textit{Bottom panel:} We show $R_\sigma(E, N_{\mathrm{config}})$ as a function of 
    $N_{\mathrm{config}}$ on the left and the relative statistical error on the right which is observed to scale as $1/\sqrt{N_{\mathrm{config}}}$, as expected.}
    \label{fig:hlt_statistics}
\end{figure}

  \textit{Stability analysis.} In order to quantify the systematic uncertainty associated with the imperfect reconstruction of the smearing kernel, we study $R_\sigma(E; \mathbf{g})$ as a function of the normalized $L_2$ norm at $\mathrm{n}=0$,
\begin{flalign}
d(\mathbf{g})=\sqrt{\frac{A_0[\mathbf{g}]}{A_0[\mathbf{0}]}}\;.
\label{eq:smalld}
\end{flalign}
We also consider the case where $\mathbf{g}$ is obtained with $\mathrm{n} \neq 0$. As shown in Fig.~\ref{fig:changelambda}, for large values of $d(\mathbf{g})$ the results obtained with different values of $\lambda$ differ significantly, indicating that the reconstructed kernels deviate substantially from the target kernel and from each other. Conversely, for sufficiently small values of $d(\mathbf{g})$, the corresponding determinations of $R_\sigma(E; \mathbf{g})$ become consistent within statistical uncertainties, which increase in this regime due to the ill-posed nature of the problem. Our best estimate for $R_\sigma(E)$ is obtained by selecting results from this statistics-dominated region (see Fig.~\ref{fig:changelambda}), namely the region of small $d(\mathbf{g})$ where stability is observed within the statistical uncertainties $\Delta_\sigma^\mathrm{stat}(E; \mathbf{g})$. In Fig.~\ref{fig:hlt_statistics}, we show the values extracted in this regime for different statistics. The fluctuations remain compatible within 2–3 standard deviations. The bottom-right panel displays the statistical error normalized to that corresponding to $N_{\mathrm{config}} = 200$, which follows the expected $1/\sqrt{N_{\mathrm{config}}}$ scaling.

\begin{figure}[!h]
\begin{center}
\includegraphics[width=0.8\textwidth]
{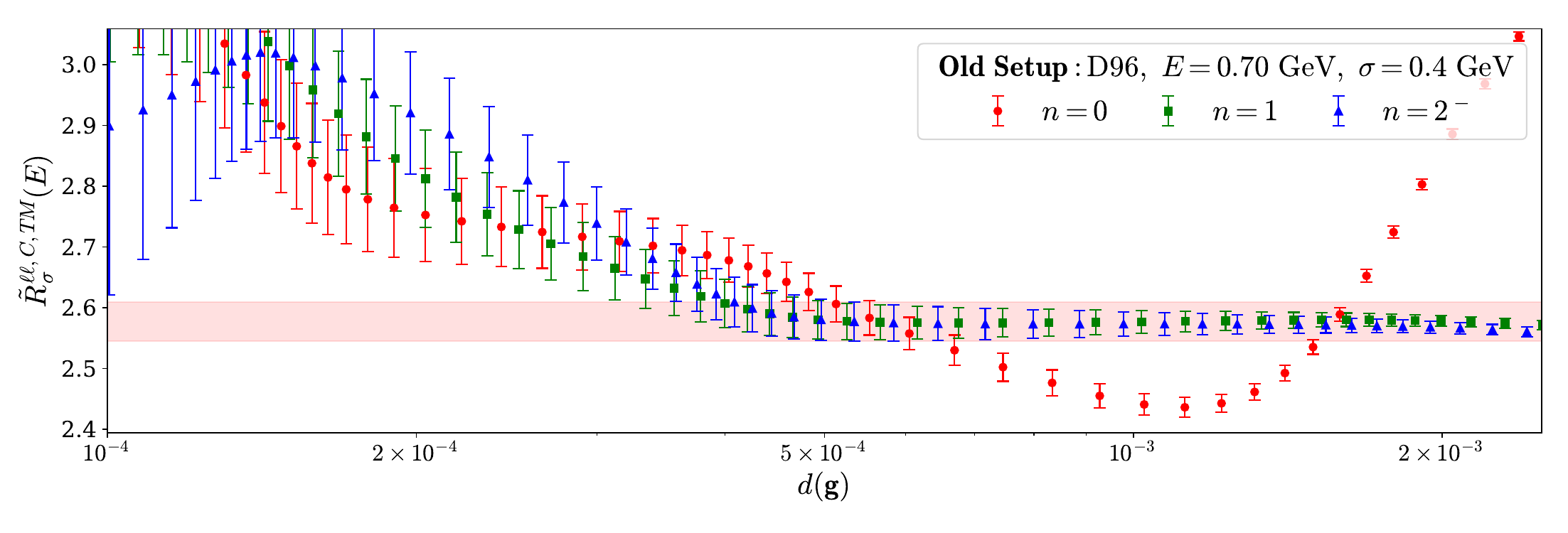}
\includegraphics[width=0.8\textwidth]
{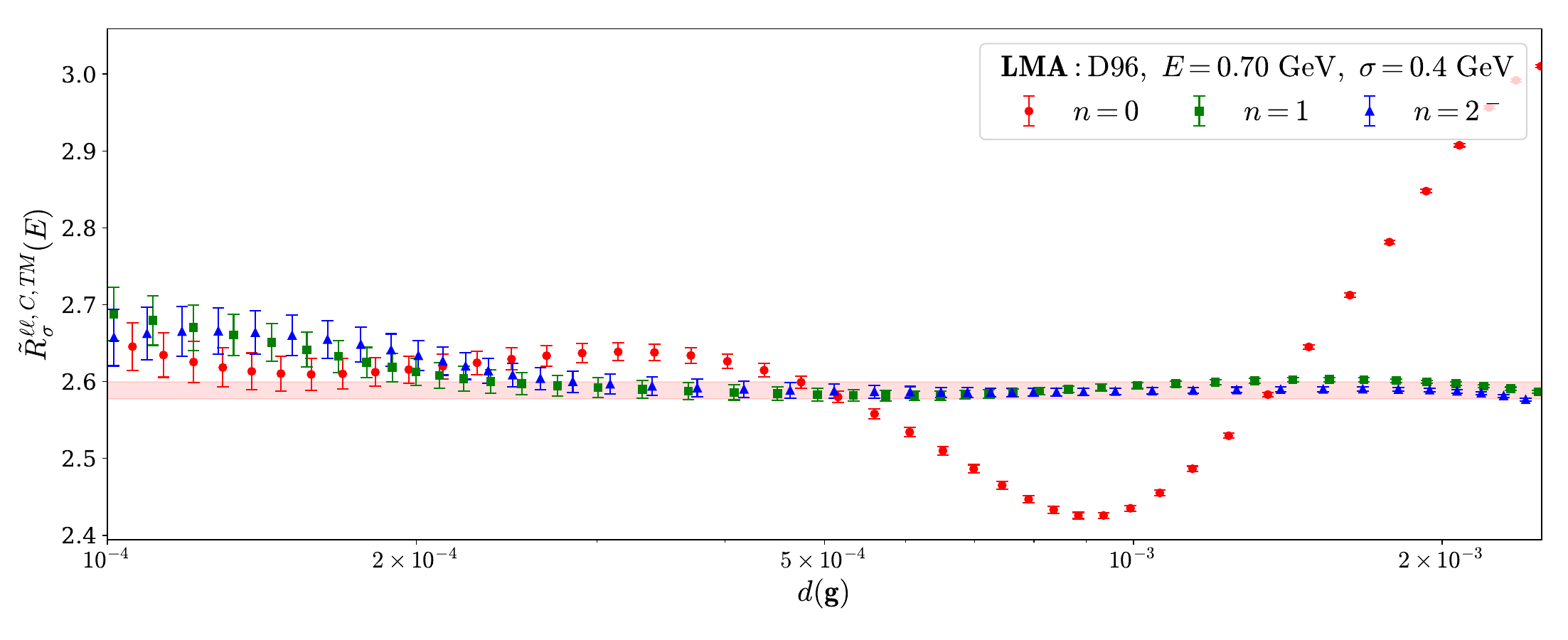}
\caption{Comparison of the stability analysis for $R^{\ell \ell}_{\sigma}(E)$ with TM regularization on the D96 ensemble at energy $E=0.70$ GeV and $\sigma=0.4$ GeV, using both the statistics and setup of Ref.~\cite{ExtendedTwistedMassCollaborationETMC:2022sta} (labelled as Old Setup in the figure) and the new setup (LMA) with increased statistics and the use of the error reduction technique.}\label{fig:stability} 
\end{center}
\end{figure}

In Fig.~\ref{fig:stability}, we compare the stability analysis for $R^{\ell \ell}_{\sigma}(E)$ with TM regularization on the D96 ensemble at $E=0.70$ GeV and $\sigma=0.4$ GeV, using both the setup of Ref.~\cite{ExtendedTwistedMassCollaborationETMC:2022sta} (old setup) and the new setup with increased statistics and the error-reduction technique (LMA). This comparison is the central result of this study and demonstrates a clear improvement in both stability and precision. In particular, the new setup enables reliable determinations of $R_\sigma(E)$ even at very small values of $d(\mathbf{g})$, where the kernel reconstruction is most accurate while the statistical uncertainties remain under control. The errors on the points are statistical, and different colours correspond to different weight functions, see Eq.~(\ref{eq:walpha}). The plot shows also the variation of results for different values of the reconstruction figure-of-merit $d(\mathbf{g})$ defined in Eq.~(\ref{eq:smalld}). Different weight functions have a significant impact on the stability of $R_\sigma(E)$ across varying $d(\mathbf{g})$, while the points obtained with $\mathrm{n}=2^-$ remain remarkably stable even at large $d(\mathbf{g})$ values. 

\begin{figure}[h]
\centering
\begin{minipage}{0.55\textwidth}
    \centering
    \includegraphics[width=0.85\textwidth]{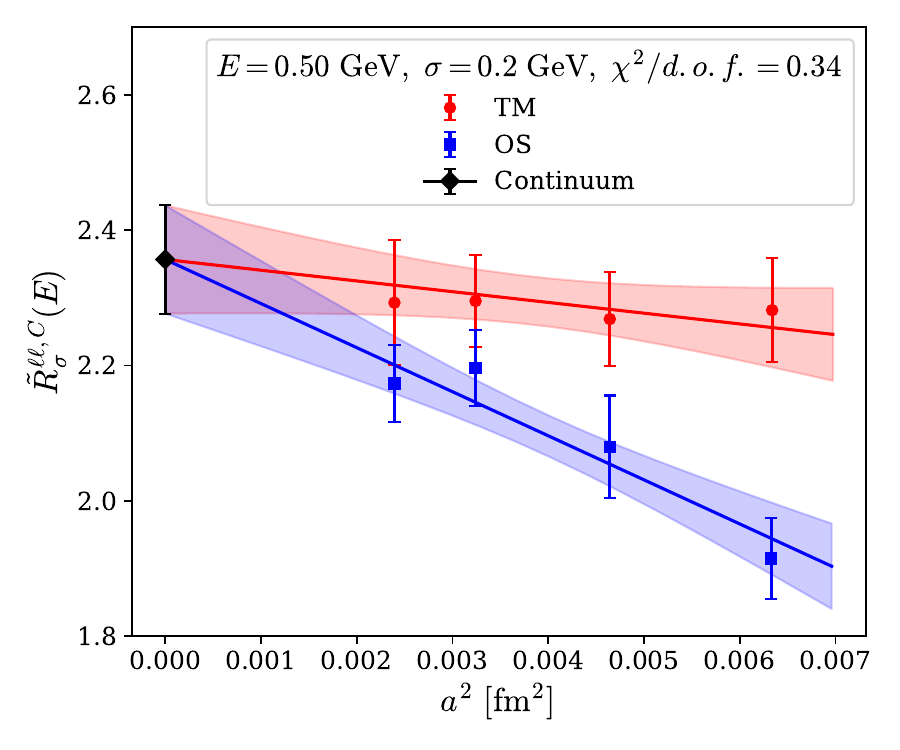}
\end{minipage}\hfill
\begin{minipage}{0.4\textwidth}
\caption{
Continuum extrapolation of $R^{\ell\ell,C}_\sigma(E)$ at $E=0.50~\mathrm{GeV}$ and $\sigma=0.2~\mathrm{GeV}$. 
Blue and red points correspond to OS and TM regularizations. 
A correlated $\chi^2$ fit is performed at fixed $E$ and $\sigma$ for $L \sim 5~\mathrm{fm}$.
}
\label{fig:continuum-limit}
\end{minipage}
\end{figure}

\textit{Continuum extrapolations.} Fig.~\ref{fig:continuum-limit} shows an example of our continuum extrapolations of $R^{\ell\ell,C}_\sigma(E)$ at $E = 0.50$~GeV and $\sigma = 0.2$~GeV. The blue and red points correspond to the OS and TM lattice regularizations, respectively. For each regularization, we perform a correlated $\chi^2$ minimization of the lattice data at fixed $E$, $\sigma$ and volume $L\sim 5$ fm.

\begin{figure}[!t]
\centering
\includegraphics[width=1\textwidth]{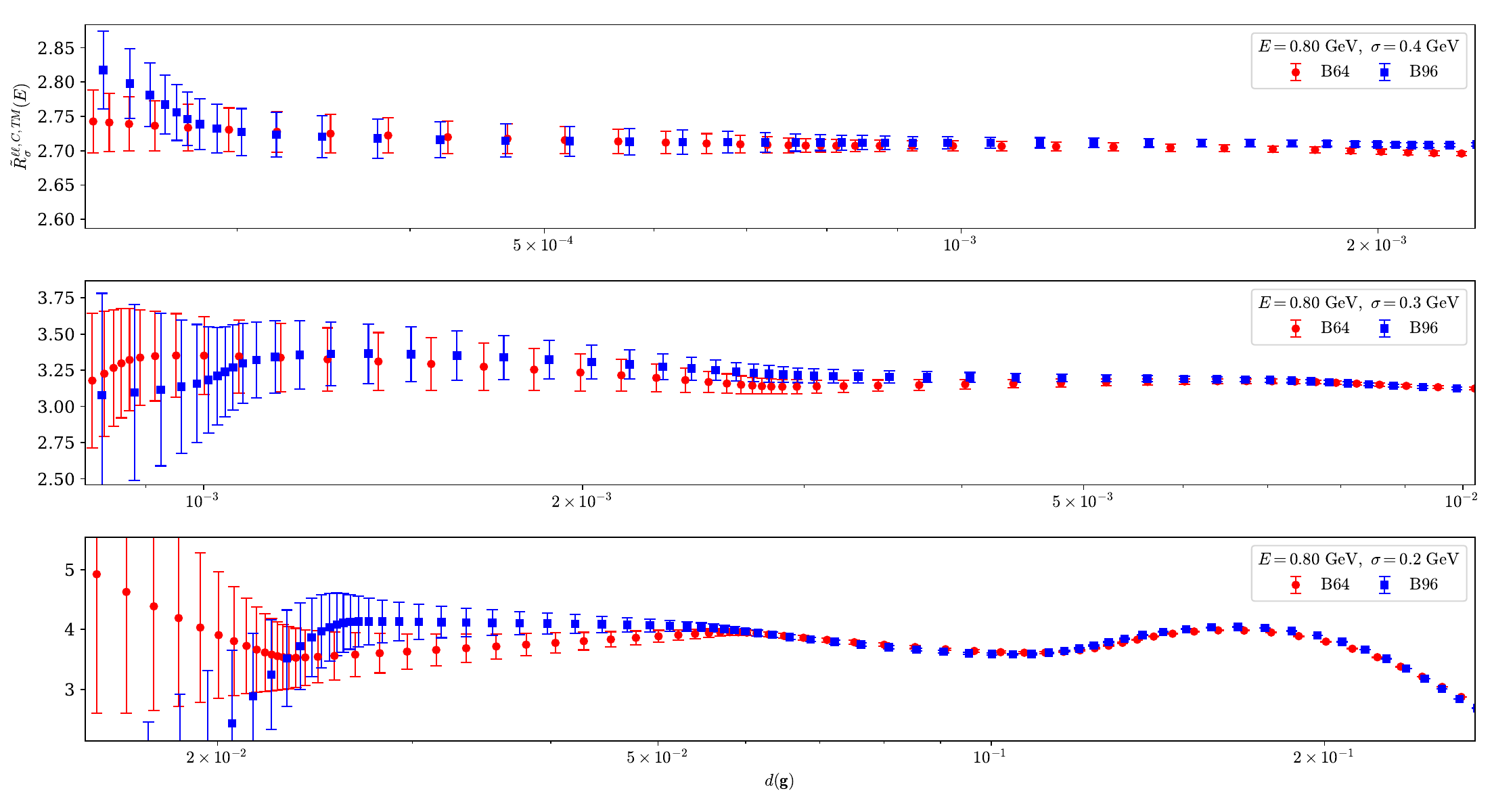} 
\caption{Stability analysis for $R^{\ell \ell}_{\sigma}(E)$ with TM regularization on the B64 and B96 ensemble at energy $E=0.80$ GeV and different values of $\sigma$.}
\label{fig:volume-effect}
\end{figure}

\textit{Volume dependence.} For smeared spectral quantities, finite-volume effects are expected to be exponentially suppressed with the lattice size $L$, as discussed in Ref.~\cite{Bulava:2021fre}. We study the volume dependence using a data-driven approach by comparing the results of $R_\sigma(E)$ extracted at $L = 5.09$~fm and $L = 7.6$~fm, and looking at the spread between these two volumes. As is evident from the stability plots shown in Fig.~\ref{fig:volume-effect}, no volume dependence is observed within uncertainties for $\sigma = 0.4$~GeV. In particular, the results obtained at different volumes lie perfectly on top of each other. We emphasize that the data points corresponding to the two volumes are derived from independent simulations and are therefore completely uncorrelated. Finite-volume effects remain small and under control even when the resolution is reduced to $\sigma = 0.2$~GeV.

\section{Conclusions}
\label{sec:results}
We are investigating the $R$-ratio by computing a smeared version  $R_\sigma(E)$, from first principles using correlation functions generated by the ETMC with $N_f = 2+1+1$ dynamical quark flavors, using four lattice spacings, different volumes, and significantly increased statistics compared to our previous study in~\cite{ExtendedTwistedMassCollaborationETMC:2022sta}. The volume dependence is studied through a data-driven approach, and we find that finite-volume effects are under control also at the smallest considered value of the smearing parameter $\sigma$.

\begin{figure}[!h]
\begin{center}
\includegraphics[width=0.9\textwidth]{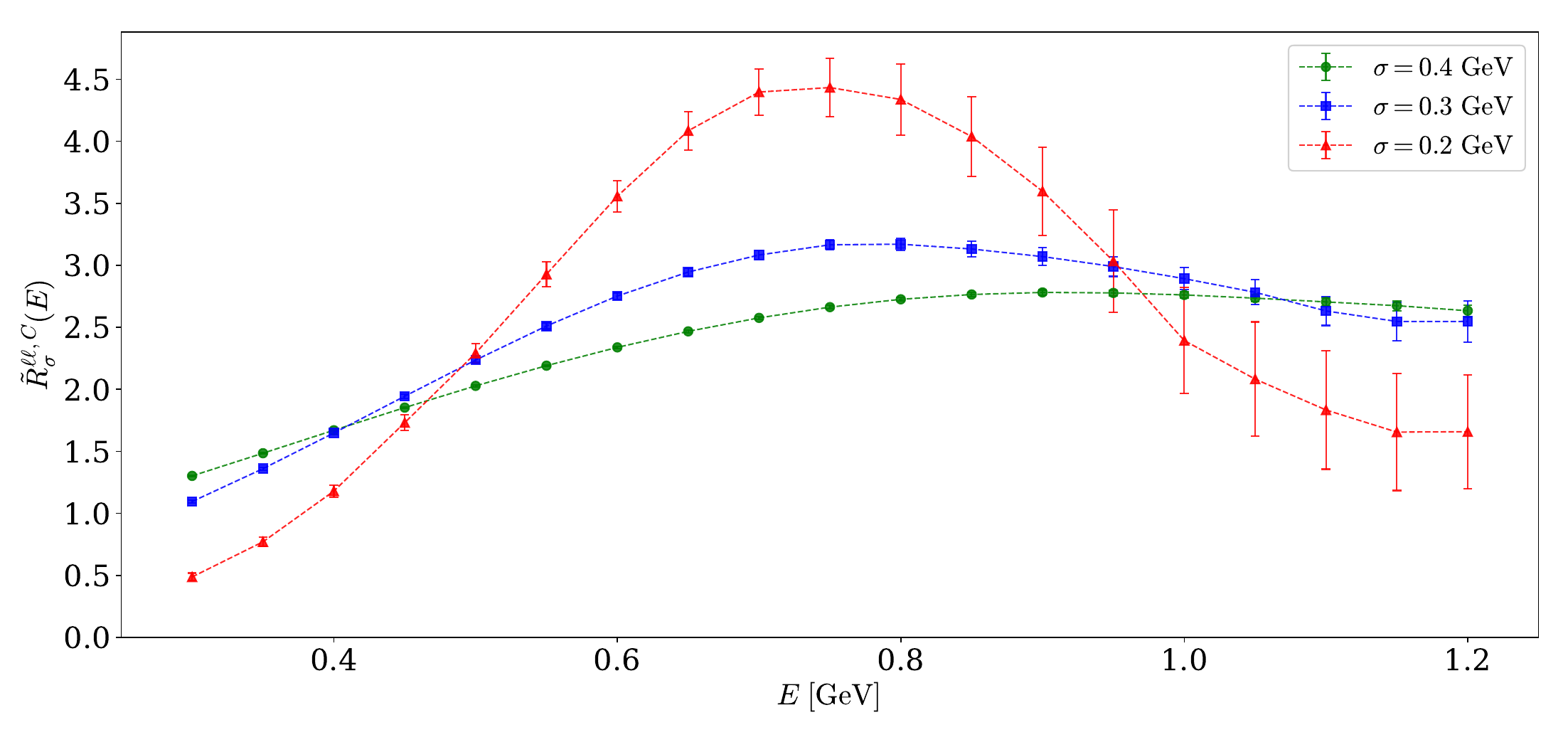}
\caption{Preliminary blinded results for the connected contribution of the $u/d$ quark mass to the smeared $R_{\sigma}(E)$, with central energies up to 1.2 GeV and $\sigma = 0.4, 0.3$ and 0.2 GeV.}\label{fig:r-ratios}
\end{center}
\end{figure}

Figure~\ref{fig:r-ratios} shows our preliminary blinded results for the connected light quark contribution to the smeared quantity $R^{\ell\ell}_{\sigma}(E)$ for three resolutions, $\sigma = 400$, 300, and 200 MeV, where $\sigma$ denotes the standard deviation of the Gaussian smearing kernel. By improving the statistical precision of the light vector–vector connected two-point correlators using the LMA technique, we reduce the relative uncertainty on $R^{\ell\ell}_{\sigma}(E)$ to about 1\% at $E \simeq 770$ MeV for $\sigma = 400$ MeV. This increased precision allows us to clearly resolve the $\rho$ resonance even for smearing widths as small as 200 MeV.

We plan to complete the full determination of $R_\sigma(E)$ by including the connected $\ell\ell$, $ss$, $cc$, contributions, as well as the disconnected contributions in isoQCD, with all relevant systematic uncertainties under control, such as kernel reconstruction, finite-volume effects, and continuum extrapolation. In the future we also plan to compute from first principles the missing QED and strong isospin-breaking corrections.

\section*{Acknowledgements}
A.E. and S.B. acknowledge support from EXCELLENCE/0524/0017 (MuonHVP) and EXCELLENCE/0524/0459 (IMAGE-N), co-financed by the European Regional Development Fund and the Republic of Cyprus via the Research and Innovation Foundation under the Cohesion Policy Programme ``THALIA 2021--2027''. 
A.E., R.F., F.M. and N.T. are supported by the Italian Ministry of University and Research (MUR) under PRIN 2022 project ``Non-perturbative aspects of fundamental interactions, in the Standard Model and beyond'' (PNRR-M4C2-I1.1-PE2, NextGenerationEU, F53D23001480006). 
F.S. is supported by ICSC – Centro Nazionale di Ricerca in High Performance Computing, Big Data and Quantum Computing, funded by European Union -NextGenerationEU and by Italian Ministry of University and Research (MUR) project FIS 00001556.
This work was supported by the Swiss National Science Foundation (SNSF) through the grants \href{https://data.snf.ch/grants/grant/208222}{208222} and \href{https://data.snf.ch/grants/grant/10003675}{10003675}.
We acknowledge CINECA and the EuroHPC Joint Undertaking for granting access to the Leonardo Supercomputer. Computing time on Leonardo Booster was provided through the Extreme Scale Access Call (grant EHPC-EXT-2024E01-027), with additional GPU resources under the INFN-LQCD123 initiative.
We acknowledge the Swiss National Supercomputing Centre (CSCS) access to Alps through the Chronos programme under project ID CH15.
The authors acknowledge the Gauss Centre for Supercomputing e.V. for funding this project by providing computing time on the GCS  Supercomputer JUWELS \cite{JUWELS} at J\"ulich Supercomputing Centre (JSC).
\bibliographystyle{JHEP}
\bibliography{biblio}

\end{document}